\documentclass[12pt,preprint]{aastex}

\shorttitle{Sky Brightness at Cerro Tololo}
\shortauthors{Krisciunas et al.} 

\begin{document}

\title{The First Three Rungs of the Cosmological Distance Ladder}
\author{
Kevin Krisciunas,\altaffilmark{1}
Erika DeBenedictis,\altaffilmark{2}
Jeremy Steeger,\altaffilmark{3}
Agnes Bischoff-Kim,\altaffilmark{4}
Gil Tabak,\altaffilmark{5}
and Kanika Pasricha\altaffilmark{6}
}

\altaffiltext{1}{George P. and Cynthia Woods Mitchell Institute for Fundamental 
Physics \& Astronomy, Texas A. \& M. University, Department of Physics and Astronomy,
  4242 TAMU, College Station, TX 77843; {krisciunas@physics.tamu.edu} }
\altaffiltext{2}{MSC \#227, Caltech, Pasadena, CA 91126; {edebened@caltech.edu} }
\altaffiltext{3}{3 Ames Street, Box 62, Cambridge, MA 02142; {jsteeger@mit.edu} }
\altaffiltext{4}{Department of Chemistry, Physics and Astronomy,
Georgia College \& State University, Campus Box 82, Milledgeville, GA 31061;
{agnes.kim@gcsu.edu} }
\altaffiltext{5}{6530 El Colegio Road, Box \#3101, Santa Barbara, CA 93106; 
{tabak.gil@gmail.com} }
\altaffiltext{6}{2269 Frist Campus Center, Princeton University, Princeton, NJ
08544; {pasricha@princeton.edu} }

\begin{abstract} 
It is straightforward to determine the size of the Earth and the 
distance to the Moon without making use of a telescope.  The methods have
been known since the 3rd century BC.  However, few amateur or professional astronomers
have worked this out from data they themselves have taken.  Here we use
a {\em gnomon} to determine the latitude and longitude of South Bend,
Indiana, and College Station, Texas, and determine a value of the radius
of the Earth of 6290 km, only 1.4 percent smaller than the true value.
We use the method of Aristarchus and the size of the Earth's shadow during
the lunar eclipse of 2011 June 15 to derive an estimate of the distance
to the Moon (62.3 R$_{\earth}$), some 3.3 percent greater than the true mean
value. We use measurements of the angular motion 
of the Moon against the background stars over the course of two nights, using 
a simple cross staff device, to estimate the Moon's
distance at perigee and apogee.  Finally, we use simultaneous CCD 
observations of asteroid 1996 HW1 obtained with small telescopes in Socorro, New Mexico, 
and Ojai, California, to derive a value of the Astronomical Unit of (1.59 $\pm$ 0.19) $\times$ 
10$^8$ km, about 6 percent too large.  The data and methods presented here 
can easily become part of a beginning astronomy lab class.
\end{abstract}


\section{Introduction}


As scientists and as educators we ask ourselves, ``How do we know what 
we know?''  If we had to derive everything we used from scratch, we would
not have the time to make much progress.  But it is worthwhile to derive 
{\em some} fundamental parameters that we use in our area of expertise.
In 1987 MIT physicist Philip Morrison hosted
a television series called ``Ring of Truth: an inquiry into how we know what
we know.''  One purpose of the show was to demonstrate simple measurements.
In the episode called ``Mapping'' he measured the circumference
of the Earth with the ``van of Eratosthenes''.\footnote[7]{See
http://www.viddler.com/explore/jacksmernov/videos/6.  Of course, they are
referring to the Greek astronomer Eratosthenes (ca. 276-195 BC) who obtained
the first estimate of the Earth's circumference from observations of the
Sun's elevation on the summer soltice from Alexandria and a town 7 degrees
to the south.}  He and his crew went to
the north edge of Kansas and measured the elevation angle of the bright
southern star Antares as it transited the meridian.  Then they drove 370 miles 
down Highway 183 to the south edge of Kansas and measured Antares again.  
They found that it was 5 degrees higher at the second location.  The distance
corresponded to 1/72 of a circle, so the implied circumference of the
Earth was 26,600 miles, which is a bit more than 6 percent greater than the 
correct value (24,901 miles).

The cosmological distance ladder is a sequence of steps used by astronomers
to derive distances within the solar system, throughout the Galaxy,
and beyond to the farthest galaxies detectable \citep{Row85}.  It hinges
on simple geometry and the principles of surveying.  For example, using the
positions of two observers separated by some baseline on the Earth, we
can determine the distance to the Moon, a nearby planet or an asteroid.  

In this paper we determine the sizes of the first three rungs
of the cosmological distance ladder.  These are: 1) the radius of the 
Earth; 2) the distance to the Moon; and 3) the distance to the Sun (i.e.
the size of the Astronomical Unit, or AU).  The first two can be accomplished
without a telescope.  The third was attempted by various pre-telescopic
astronomers such as Tycho Brahe in 1582 \citep{Gin_Voe98}, but was not accomplished
until 1672 by Gian Domenico Cassini and John Flamsteed \citep{Van95}, who made
observations of Mars when it was prominently visible in the middle of the night
and therefore about as close to the Earth as it gets.  (Mars is roughly 0.6 
AU distant at such a time.  When it is on the other side of the Sun it is 2.6 AU
distant.)    Measuring the size of the AU requires 
telescopic measurements or distances to nearby planets or asteroids determined
with radar.

Once the scale
of the solar system is known, we can determine the distances to nearby stars
via trigonometric stellar parallaxes if our positional measurements are good 
to better than 0.1 arc second.  Then, classically, we determine the distance
to the Hyades star cluster \citep[][p. 48]{Row85} and tie other star
clusters to the Hyades distance.  With the discovery of certain standard
candles and standardizable candles in star clusters (e.g. RR Lyrae stars
and Cepheids), we can calibrate the mean intrinsic brightness of these pulsating
stars.  They are useful for distance determinations throughout our Galaxy.
With the Hubble Space Telescope we can determine distances to other
galaxies using the Cepheid period-luminosity relation as far as 25 million parsecs
\citep{Rie_etal10}.  (1 parsec equals 3.086 $\times$ 10$^{13}$ km, or 206265 AU.) 
If we observe Type Ia supernovae in some galaxies whose distances are known
via Cepheids, we can calibrate the intrinsic brightness of these supernovae.  Since 
Type Ia supernovae are typically 4 billion times brighter than the Sun at maximum
brightness, they can be used to determine distances halfway across the observable
universe with a 4-m class telescope.

The methodology of this paper is to use only our own data, if possible.  We
make minimal use of the {\em Astronomical Almanac} and minimal use of telescopes.
Many of the results here can be obtained independently by students using very 
unsophisticated and inexpensive equipment.  

\section{Determining one's position on the Earth}

In Fig. \ref{triangle} we show the astronomical triangle.  The northern sky turns
around the North Celestial Pole (NCP), near the direction of the star
Polaris.  The NCP is $\phi$ degrees above the horizon.  
$\phi$ is the latitude.  The azimuth ($A$) is measured clockwise around the horizon
and is equal to 0$^{\circ}$ deg at the north point on the horizon, 90$^{\circ}$ deg at the
east point on the horizon, etc.  The hour angle $t$ divided by 15 deg/hr is the number
of hours that an object is west of the celestial merdian.  $t$ is negative for
objects in the eastern sky.

The declination ($\delta$) of a celestial object is the number of degrees the
object is north or south of the celestial equator.  
If we happen to know the declination of the Sun and we determine how
high it is above the horizon when it transits the {\em celestial 
meridian},\footnote[8]{The celestial meridian is the imaginary line in the sky
that separates the east half from the west half.  For an observer in the northern
hemisphere the meridian extends from the north point on the horizon through the North
Celestial Pole, through the zenith and down to the south point
on the horizon.  An object such as the Sun is at its highest point above the
horizon when it crosses the celestial meridian.  At that moment the local apparent
solar time is, by definition, exactly 12 noon.} we
can determine our latitude.  At local apparent noontime the elevation angle 
of the Sun will be:

\begin{equation} 
h_{max} \; = \; 90^{\circ} \; - \; \phi \; + \; \delta \; .
\end{equation}

\parindent = 0 mm

We are concerned here with the Sun and Moon as observed at mid-northern 
latitudes.  These objects transit the celestial meridian between the
zenith and the south point on the horizon.

\parindent = 9 mm

To determine the elevation angle of the Sun we can set up a vertical pointed
stick or {\em gnomon}.  Even better is a vertical stick with a small sphere
at the top, like that shown in Fig. \ref{gnomon}.  It is easier to measure
the center of the elliptical shadow of the sphere on the ground than it is
to measure the end of the darker part of the shadow of a vertical pointed 
stick.\footnote[9]{If a vertical pointed stick is used, this will give us
the elevation angle of the upper limb of the Sun.  To find the elevation
angle of the center of the Sun then requires subtracting the angular radius
of the Sun.}

The Sun's declination ranges from $-\epsilon$ to $+\epsilon$ over the course
of the year.  $\epsilon$ is the {\em obliquity of the ecliptic}, the tilt of
the Earth's axis of rotation to the plane of its orbit.  (The ecliptic is
the apparent path of the Sun through the constellations of the zodiac owing
to the Earth's orbit around the Sun.) On the first day of summer we have:

\begin{equation}
h_{max} (\rm{Jun} \;  21) \; = \; 90^{\circ} \; - \; \phi \; + \; \epsilon \; \; .
\end{equation}

\parindent = 0 mm

And on December 21st we have:

\begin{equation}
h_{max} (\rm{Dec} \; 21) \; = \; 90^{\circ} \; - \; \phi \; - \; \epsilon \; \; .
\end{equation}

Eq. 2 plus Eq. 3 gives:

\begin{equation}
\phi \; = \; 90^{\circ} \; - \; [ h_{max} (\rm{Jun} \; 21) \; + \; h_{max} (Dec \; 21) ] / 2  \; \; .
\end{equation}

And Eq. 2 minus Eq. 3 gives us:

\begin{equation}
\epsilon \; = \; [ h_{max} (\rm{Jun} \; 21) \; - \; h_{max} (Dec \; 21) ] / 2  \; \;  .
\end{equation}

In other words, if we determine the maximum elevation angle of
the Sun on June 21st and December 21st, the average gives us 90$^\circ -\phi$,
and half the difference gives us $\epsilon$.

\parindent = 9 mm

In Table \ref{gnomon_data} we give the raw data from two solstice experiments
carried out in College Station, Texas.

In Fig. \ref{gnomon_xy} we show the X-Y positions of the end of the
gnomon's shadow for three key times of the year.  Note that when the
declination of the Sun is negative the positions of the end of the shadow
on the ground trace out an upward pointing curve.  On the first day of
spring (March 20 or 21) or the first day of autumn (about September 23)
the points delineate a straight line.  When the declination of the Sun is positive
the points trace out a downward curving locus.  A graph of this type
allows one to determine which points, if any, have been mismeasured.  If a student
invents data for such an experiment, such a graph allows one to detect
such fraud easily.

Plots of the data from Table \ref{gnomon_data} are shown in Fig. \ref{shadow}.
We have converted the time values to the number of minutes since local
apparent noontime (when $h = h_{max}$).\footnote[10]{Apparent solar time
relates to the hour angle of the actual visible Sun in the sky.  Due to the
tilt of the Earth's axis of rotation to the plane of its orbit, and due
to the ellipticity of the Earth's orbit, apparent solar time ranges from
14 minutes behind to 16 minutes ahead of mean solar time.  Our watch time
is basically mean solar time adjusted (by longitude difference)
to the nearest 15 degree line of longitude west of Greenwich, England.
For example, Central Standard Time is for locations that are about 90$^{\circ}$
longitude west of Greenwich, or 6 hours west.}
In the top graph we have added, for graphical purposes only, a hyperbolic fit
to the data.  In fact, there is no simple function that fits such a data set.
The best estimate of the minimum value in the summer solstice plot
is obtained from a fourth order polynomial
fit to the data within 41 minutes of the meridian transit of the Sun.  
For the 2010 June 21 observations we used a gnomon of height $g = 632 \pm 1$ mm.
The minimum shadow length was L$_{min}$ = 85.84 mm with an average
scatter of $\pm$ 0.64 mm.  This average scatter is what is known as a 
{\em random error}, as the points scatter randomly above and below an
appropriate function fit to the data.  $h_{max}$ is equal to the arctangent 
of $g/L_{min}$, or 82$^{\circ}$ 15\farcm~9 $\pm$ 3\farcm~5.

For 2010 December  21 we used a gnomon of height $g = 550 \pm 1$ mm.
From a fourth order polynomial fit to all 26 points we find
L$_{min}$ = 753.98 $\pm$ 0.74 mm.  $h_{max}$ is found to be
36$^{\circ}$ 06\farcm~6 $\pm$ 3\farcm~4.

Eq. 4 gives $\phi$ = 30$^{\circ}$ 48\farcm~7 $\pm$ 3\farcm~5, which is 11\farcm~5 north
of the true value of 30$^{\circ}$ 37\farcm~2 from Google 
Earth.  (One arc minute
of latitude equals one nautical mile, or 1852 m.)  Eq. 5 gives
$\epsilon$ = 23$^{\circ}$ 04\farcm~7 $\pm$ 3\farcm~5, which is several standard deviations
smaller than the true value of 23$^{\circ}$ 26\farcm~2.  These values show
what kind of uncertainty can be obtained with careful observations.  

So far we have used no information from the {\em Astronomical Almanac}.  However, 
to determine our {\em longitude} requires that we know the amount of time
that the apparent Sun is ahead or behind the mean Sun.  This difference is
known as the {\em equation of time}.  It ranges from $-$14.2 minutes to +16.4
minutes over the course of the year.  

From four gnomon experiments carried out in College Station, Texas, 
on 2006 November 12, 2007 September 24, and the two solstices mentioned 
above we found that the mean Sun transits the meridian at 12:24:26 PM standard time,
with an uncertainty of $\pm$72 sec.  Since each degree of longitude
corresponds to 4 minutes of time, our longitude is 96$^{\circ}$ 06\farcm~5 $\pm$ 18\farcm~1.
We are within one standard deviation of the true value of 96$^{\circ}$ 20\farcm~4.

\section{The circumference of the Earth}

On 2006 September 3 we carried out a gnomon experiment in South Bend, Indiana,
on the campus of the University of Notre Dame.  For this we needed a value of the
declination of the Sun from the {\em Astronomical Almanac}.  Our location
was found to be latitude 41$^{\circ}$ 51\farcm~5, longitude 85$^{\circ}$ 50\farcm~5.  

In late October of 2007 we drove from South Bend through Bella Vista, Arkansas,
and on to College Station, Texas, keeping track of our route, mileage, and the
length of any side trips.  After subtracting the side trips, the
elapsed distance on the odometer was 1248 statute miles.  For a section of highway
through central Illinois we noted that 96.8 miles on the odometer corresponded
to 98.0 miles according to the highway markers.  The implication is that our
odometer exhibited a systematic error.  On a subsequent occasion we drove the
final 92 percent of the exact same route, but with a different car, and the
odometer mileage was 40 miles less (about 4 percent).  We conclude that an 
odometer cannot be explicitly trusted.  We adjust the original mileage to 
1248 $\times \frac{98.0}{96.8}$ = 1263 miles.

Next, we used a ``map tool'' (also known as an {\em opisometer}) and traced our
route on a map from the 1980 edition of {\em The New International Atlas} by 
Rand McNally, and determined that
if we could have traveled along a great circle route from South Bend to College
Station, the direct route had a length that was  0.744 of the length of the
route we actually drove.  (This means that we did not rely on the stated scale
of the map.)  Thus, the great circle distance from one place to the other was 
940 statute miles.

Consider two locations on the Earth with (latitude, longitude) =  ($\phi _1$, $\lambda _1$)
and ($\phi _2$, $\lambda _2$), respectively.  The length of the great circle arc 
between them ($\rho$) can easily be obtained from the law of cosines of
spherical trigonometry:

\begin{equation}
\rm{cos} (\rho) \; = \; \rm{sin}(\phi _1) \; \rm{sin}(\phi _2) + 
\rm{cos}(\phi_ 1) \; \rm{cos} (\phi _2) \; \rm{cos} (\lambda _2 - \lambda _1) \; .
\end{equation}

Given the latitudes and longitudes determined by gnomon and clock at both locations,
the great circle arc was 13.78 deg.  Thus, our estimate of the circumference of
the Earth is 940 $\times \frac{360}{13.78}$ = 24557 miles, which is 1.4 percent
smaller than the true value.  Our value for the radius of the Earth is 6290 km.
The true value is 6378 km.  Given the simplicity of the tools involved, we
feel that we have achieved an accuracy far better than our expectations.

For comparison we refer the reader to \citet{Bek_etal11}, which describes projects
``Eratosthenes 2009'' and ``Eratosthenes 2010 America''.  More than 15,000 students
at more than 200 schools determined the radius of the Earth using this method.
They obtained 6290 km in 2009 and 6375 km in 2010.

\section{The distance to the Moon}

\subsection{Using the method of Aristarchus}

The method used above to determine the circumference of the Earth is in
principle the same as that used by Eratosthenes.  It turns out that
it was a generation {\em before} Eratosthenes that Aristarchus of Samos
(ca. 310-230 BC) determined the distance to the Moon in terms of the
radius of the Earth.  Inherent in both methods is the acceptance of the
idea that the Earth is, for all intents and purposes, spherical.

Aristarchus cleverly deduced that we can determine the distance to the
Moon from the geometry of a lunar eclipse.  See \citet[][pp. 68-73]{Eva98}
and Fig. \ref{lunar_ecl}.
A lunar eclipse can only take place when the Moon is opposite the Sun.
The shadow of the (spherical) Earth is circular in any plane perpendicular
to the Earth-Sun line.  Consider the
triangle delineated by points A, C, and H.  AC is the radius of the
Earth, R$_{\earth}$.  CH is the distance from the center of the Earth
to the Moon, d$_{Moon}$.  The {\em horizontal parallax} of the Moon,
P$_M$, is given by:

\begin{equation}
\rm{sin} (P_M) \; = \; \frac{R_{\earth}}{d_{Moon}} \; .
\end{equation}

\parindent = 0 mm

Thus, the distance to the Moon in Earth radii is 1/sin(P$_M$).
In Fig. \ref{lunar_ecl} P$_S$ is the horizontal parallax of the Sun.  
$\sigma$ is the angular radius of the Sun.  $\tau$ is the angular
radius of the Earth's shadow at the distance of the Moon.  For
$\bigtriangleup$XCH it is obvious that $\angle$XCH + P$_S$ + P$_M$ = 180$^{\circ}$.
In other words, 180$^{\circ} - \angle$XCH = P$_S$ + P$_M$.  Also, 
$\sigma$ + $\angle$XCH + $\tau$ = 180$^{\circ}$, which means that 
180$^{\circ} - \angle$XCH = $\sigma$ + $\tau$.  It follows that:

\begin{equation}
P_S \; + \; P_M \; = \; \sigma \; + \; \tau \; . 
\end{equation}

\parindent = 9 mm

Aristarchus believed that the Sun was 19 times more distant than the Moon.
The true ratio is closer to 400.  Our value for the radius of the Earth and
our value of the distance to the Sun given in \S 5 below means that the solar
parallax P$_S \approx$  0\farcm 14.  

Aristarchus knew that the Moon and Sun approximately had the same
angular diameter because the duration of a total solar eclipse is never
more than a few minutes.  He took the angular diameter of the Moon to
be 0.5 deg.  In the {\em Almagest}, section V 14, Ptolemy states that
the Earth's shadow at the distance of the Moon is 2.6 times the angular
diameter of the Moon \citep[][p. 254]{Too84}.  We can use
two actual measurements 
of the Sun made on 2010 May 1 and 2011 June 6 to provide an angular diameter
of 30\farcm 8 $\pm$ 0.9, so $\sigma \approx$ 15\farcm 4.  

In a previous
paper \citep{Kri10} we presented measures of the Moon's angular diameter
from sightings of the Moon through a 6.2 mm hole in a piece of cardboard
that slides up and down a yardstick.  Our mean value
for the angular diameter of the Moon was 31\farcm 18.
Data in that paper and subsequent data yield a time of lunar
perigee of 2011 May 14.70 UT, with a perigee-to-perigee period (i.e.
the {\em anomalistic month}) of 27.4992 $\pm$ 0.0518 d, which is within
1-$\sigma$ of the correct value of 27.55455 d.  Based on our non-telescopic
data, the lunar eclipse of 2011 June 15 occurred about 4.6 days after perigee, 
so the Moon's angular size would have been a bit larger than the mean value.

Using six images obtained from www.slooh.com,
we determined that the Earth's shadow was 2.56 $\pm$ 0.03 times the 
angular diameter of the full Moon during the lunar eclipse of 2011 June 15.  
Thus, $\tau$ = 0.5 $\times$ 31.18 $\times$
2.56 $\approx$ 39\farcm 19.  It follows that P$_M$ =  15\farcm 4 + 39\farcm 19 $-$
0\farcm 14 $\approx$ 55\farcm 17. Eq. 7 gives a lunar distance of 
62.3 R$_{\earth}$ with a random error of at least $\pm$ 0.7 R$_{\earth}$.
The true mean distance is 60.27 R$_{\earth}$.

\subsection{Using the motion of Moon against the stars}

We have seen that the geometry of a lunar eclipse allows a determination
of the distance to the Moon.  Hipparchus (ca. 140 BC) used observations
of the {\em solar} eclipse of 14 March 190 BC to derive a distance to the
Moon between 71 and 83 R$_ {\earth}$ \citep {Too81}.  But can one determine
the Moon's distance without the use of an eclipse?  Simultaneous observations
of the Moon's position from two widely separated places on the Earth
would suffice.  \citet[][pp. 200-202]{Sma77} shows how this can be
done from two observatories widely separated in latitude but on the same
meridian of longitude.

Observations of the Moon from the {\em same} location 
over the course of a night can also be used to determine the distance to 
the Moon.  In
this case one also uses a large fraction of the Earth's diameter as a
baseline.  The complication is that while we wait for the Earth
to rotate, the Moon is also orbiting the Earth.  

The Moon's orbital period with respect to the background stars is 27.32 d.  
Thus, {\em on average}, the Moon moves
0.55 deg east per hour against the background of stars as viewed by a
hypothetical observer at the Earth's center.  However, because of
the parallax of the Moon, the observed motion for someone situated on the
surface of the Earth is roughly 1/6 deg per hour less \citep[][pp. 252-254]{Eva98}. 

We need the geocentric vs. topocentric
shifts in right ascension and declination, respectively, of the Moon owing to
our location on the surface of the Earth.  The {\em topocentric} coordinate
system has the observer, standing on the Earth's surface, at the origin.
Let $\Delta \alpha$ = 
$\alpha ^{\prime} - \alpha$ = $t - t^{\prime}$. (Right ascension increases to
the east, while hour angle increases to the west.) Then Eq. 35 of \citet{Sma77} gives

\begin{equation}
\rm {tan} \; \Delta \alpha \; = \; - \left ( \frac{R_{\earth}}{d_ {Moon}}\right ) \; \frac {\rm{sin} \; t \; \rm{cos} \; 
\phi^{\prime}} {\rm{cos} \; \delta \; - \; \left (\frac{R_ {\earth}}{d_ {Moon}}\right ) \; \rm {cos} \; t \;
 \rm{cos} \; \phi^{\prime} } \; \; .
\end{equation}

\parindent = 0 mm

Letting $\delta^{\prime} = \delta + \Delta \delta$ and T = tan$\Delta \delta$, Eq. 38
of \citet {Sma77} is

\begin{equation}
\frac{\rm{tan} \; \delta \; + \; T}{1 \; - \; T \; \rm{tan} \; \delta} \; = \; 
\frac{\rm{cos} \; t^{\prime} \left [\rm{sin}\; \delta \; - \; \left (\frac{R_ {\earth}}{d_ {Moon}}\right ) \; \rm {sin} 
\; \phi^{\prime}\right ] }
{\rm{cos} \; \delta \; \rm{cos} \; t \; - \; \left (\frac{R_ {\earth}}{d_ {Moon}}\right ) \; \rm{cos} \; \phi^{\prime}} \; \; .
\end{equation}

Here the primed parameters are the topocentric values of
declination, hour angle, and latitude, while unprimed values are for a 
hypothetical observer at the Earth's center.  Eqs. 9 and 10 cannot
be easily inverted to give d$_{Moon}$/R$_{\earth}$. But since these
equations contain this ratio, we now show how single-site observations 
can be used to obtain the distance to the Moon.  What follows is primarily a
demonstration of method.

\parindent = 9 mm

In Fig. \ref{moon_rate} we show the rate at which the Moon's apparent
position (for an observer situated in College Station, Texas) varies
as a function of hour angle for a number of occasions over nearly
the full range of distance of the Moon from the Earth.  Not surprisingly,
the Moon moves against the background of stars more slowly on a night when it is
near apogee than when it is near perigee.  

In Fig. \ref{cross_staff} we show a simple cross staff.  The pattern for the
device that slides up and down the yardstick was obtained from the University
of Washington.\footnote[11]{http://www.astro.washington.edu/courses/labs/clearinghouse/labs/Skywatch/angles.html}
With such a device we can derive the angular separation of two objects in
the sky.  For two objects with known
right ascensions and declinations $\rho$ is given by an expression 
similar to Eq. 6:

\begin{equation}
\rm{cos} (\rho) \; = \; \rm{sin}(\delta _1) \; \rm{sin}(\delta _2) + 
\rm{cos}(\delta_ 1) \; \rm{cos} (\delta _2) \; \rm{cos} (\alpha _2 - \alpha _1) \; .
\end{equation}

If the Moon's angular separation from two stars of known RA and DEC is
determined, then it is possible to determine the RA and DEC
of the Moon.  The easiest way to envision this is to take
a star chart and a compass and draw portions of two circles of
different radius centered at the locations of the two particular
stars.  One obtains two numerical solutions of the Moon's position, one 
on each side of the great circle arc joining the two stars.  One
of those possibilities can easily be eliminated.
Once an approximate location of the Moon's position is found,
one can use a computer program to search a box that covers a range
of RA and DEC to find the celestial location that is the specified
number of degrees from each of the two stars of known position.

In Table \ref{separations} we give five sets of observations made on the
night of 2011 May 15 (UT), an occasion when the Moon was close to the bright
star $\alpha$ Vir and within a day of perigee.  On this night the Moon
was waxing and 93 percent illuminated.  Thus, it was visible almost
the entire night.  One of our reference
``stars'' was Saturn.  Given the accuracy of our observations we can assume
that the position of Saturn was constant on this night.  As one can see, 
the angular separation of Saturn and $\alpha$ Vir was measured to 
better than 0.2 deg on five occasions.  Because the Moon is not a point
source, measuring the angular separation of the Moon and Saturn or the 
Moon and a star is more difficult than measuring the angular separation 
of two bright stars or one bright star and a planet.

In Table \ref{posns} we give the derived right ascensions and declinations
of the Moon from our observations of 2011 May 15.  At the end of the
night the reference stars $\delta$ Crv and $\gamma$ Hya were too low in the sky
to be seen.  The uncertainties of the right ascension and declination
of the Moon at 08:03 and 08:41 UT were derived with the
assumption of an uncertainty of $\pm$~0.3 deg for the angular
separation of Saturn vs. the Moon and $\alpha$ Vir vs. the Moon.
Because of the large uncertainty in the RA and DEC of these final
two observations, we exclude data sets four and five from further analysis.
The first, second, and third determinations of the position of the Moon
yield an apparent angular rate of 0.508 $\pm$ 0.094 deg/hr.
The corresponding range of the Moon's distance is 47.3 to 58.5 R$_{\earth}$
with a most likely value of 52.1 R$_{\earth}$.

2010 October 21 UT was the day after lunar apogee.  On this date we made a 
series of six measurements of the angular separation of the Moon vs. 
Jupiter over a 9.1 hour period. (No stars near the Moon were bright enough 
to be seen with the unaided eye given the quality of the sky.)  Thus, we 
could not use the same method to obtain multiple estimates of the RA and
DEC of the Moon at a given time.  The
separation of the Moon and Jupiter increased from 11.31 to 14.77 deg, with 
a mean rate of increase of 0.403 $\pm$ 0.028 deg/hr.  Under the assumption 
that the Moon was moving directly away from the position of Jupiter (which 
was only approximately true), this angular rate is a lower limit to the rate 
of change of position of the Moon against the background of stars. The 
implied range of lunar distance was 57.2 to 62.1 R$_{\earth}$, with a most 
likely value of 59.5 R$_{\earth}$.

If we take the 1-$\sigma$ upper limit of the rate of change of position of 
the Moon on 2011 May 15 and the 1-$\sigma$ lower limit of the rate of 
change of the lunar position on 2010 October 21, we get a conservative 
estimate of the range of the Moon's distance, namely 47 to 62 R$_{\earth}$. 
Obviously, the observations and analysis required for this method of 
determining an estimate of the Moon's distance are much more complicated 
than Aristarchus's method using the geometry of a lunar eclipse.

\section{The distance to the Sun}

In 2008 the authors of this paper were participants in the Summer Science Program
(SSP), a residential non-credit enrichment program for incoming high school juniors and
seniors.  Coauthors Krisciunas and Kim were faculty, while the other coauthors
were students. It has been the tradition at SSP to divide the students
into teams of three for the observing, but each student then writes computer
code to produce a determination of the orbital parameters of a particular
asteroid.  In Table \ref{orbit_det} we give the orbital solution of asteroid
8567 (= 1996 HW1) by one of us (J. S.)
along with the parameters obtainable from the Horizons website of the Jet
Propulsion Laboratory.\footnote[12]{ssd.jpl.nasa.gov/horizons.cgi\#top}
The code to determine the orbital parameters was debugged using observations
of Ceres and the orbital parameters of Ceres given by JPL Horizons.
We note that our orbit solution is based on observations extending over
a very small arc of the full orbit.  (The orbital period of asteroid 1996 HW1 
is 2.93 years.)

It had been an ambition at SSP for some time to observe
a near Earth object simultaneously from two sites so that its distance
could be derived, the length of the Astronomical Unit could be measured,
and the scale of the solar system be determined.  We report one
such experiment here.

On 2008 July 24 UT we took images of asteroid 1996 HW1 at  Etscorn Observatory 
of the New Mexico Institute of Mining and Technology in Socorro, New Mexico, using a
15 cm Takahashi refractor and CCD camera.  The geographic position was
latitude +34$^{\circ}$ 04$^{\prime}$ 21\farcs 7, 
longitude W 106$^{\circ}$  54$^{\prime}$ 50\farcs 1.  
A simultaneous image was obtained in Ojai, California, at 
latitude +34$^{\circ}$ 26$^{\prime}$ 04\farcs 0, 
longitude W 119$^{\circ}$  11$^{\prime}$ 22\farcs 6 using a
25 cm Meade reflector and CCD camera.  Figs. \ref{pos3}
and \ref{pos4} show the asteroid at 08:17:28 UT
on 2008 July 24, as viewed from the two sites.

The right ascensions and declinations of a number of field stars were
determined using {\bf ds9} and an image of the field obtained from the
Space Telescope Science Institute Digital Sky Survey.\footnote[13]{
http://archive.stsci.edu/cgi-bin/dss\_form}
We then determined a transformation from pixel coordinates to right ascension 
and declination using the {\sc iraf} programs {\bf ccmap} and 
{\bf cctrans}.\footnote[14]{{\sc iraf} is distributed by the National
  Optical Astronomy Observatory, which is operated by the Association
  of Universities for Research in Astronomy, Inc., under cooperative
  agreement with the National Science Foundation (NSF).}  This allowed
us to determine the right ascension and declination of the asteroid for 
each of our images.
From the New Mexico site we determined the asteroid's topocentric position to be
$\alpha$ = 21:27:07.42, $\delta$ =  +15$^{\circ}$ 53$^{\prime}$  02\farcs 77.
From the Etscorn image and the positions of the field stars, the root-mean-square
errors in right ascension and declination were $\pm$ 0.31 and $\pm$ 0.36 arcsec, respectively.
Our image obtained in California just barely has the asteroid in the frame,
and the guiding was not as good.  Still, we find a topocentric position of
$\alpha$ = 21:27:07.77, $\delta$ =  +15$^{\circ}$ 53$^{\prime}$  02\farcs 25.
The RMS errors of the field star positions were $\pm$ 0.52 and  $\pm$ 0.44
arcsec, respectively, for right ascension and declination. 

The effect on right ascension and declination due to parallax and
the finite size of the Earth can be calculated much more easily than
using Eqs. 9 and 10 (above) if the planet or asteroid under consideration
has a distance $d$ considerably greater than the distance to the Moon.
\citet[][pp. 209-210]{Sma77} gives these relevant equations:

\begin{equation}
\Delta \alpha \; = \; - \left (\frac{R_{\earth}}{d}\right ) \; \rm{sin} \; t \; \rm{cos} \; \phi^{\prime} 
\;  \rm{sec} \;\delta \; .
\end{equation}

\begin{equation}
\Delta \delta \; = \; - \left (\frac{R_{\earth}}{d}\right) (\rm{sin} \; \phi^{\prime} \; \rm{cos} \; \delta 
\; - \; \rm{cos} \; \phi^{\prime} \; \rm{sin} \; \delta \; \rm {cos} \; t) \; .
\end{equation}

\parindent = 0 mm

The values on the left hand sides of these equations are measured in radians.

\parindent = 9 mm

Given that the latitudes of the New Mexico and California sites were almost
the same, we have almost no leverage to use Eq. 13 to determine the distance
to the asteroid.  We limit ourselves to a consideration of the effect
of parallax on the right ascensions.  Let the New Mexico site be ``position 1''
and the California site be ``position 2''.  Consider the seconds part of the
observed right ascension of the asteroid.  There exist corrections to the
right ascension such that 7.77 $- \; c_2$ = 7.42 $- \; c_1$.  The corrections
$c_i$ adjust the observed right ascensions to what would be observed by
a hypothetical observer at the center of the Earth.  So, 0.35 seconds
of time equals the difference of the parallactic corrections.  Since
one second of time in right ascension equals 15 cos $\delta$ arc seconds,
we have 5\farcs 05 $\pm$ 0.61 = $c_2 - c_1$.  The uncertainty comes
from the square root of the sum of squares of the uncertainties 
of the pixel to RA/DEC transformations from {\bf ccmap} in {\sc iraf}.  

At the New Mexico site at the time Fig. \ref{pos3} was taken the hour angle
of the asteroid was $-$1.899 deg, while at the California site the
hour angle of the asteroid was $-$14.176 deg when Fig. \ref{pos4}
was taken.  Using Eq. 12 we obtain a distance to the asteroid of:

\begin{equation}
d \; = \; \frac{R_ {\earth} (0.210010 - 0.028541)}{ \left (\frac{5.05 \pm 0.061} {206265}\right ) } \; .
\end{equation}

\parindent = 0 mm

The reader will know that the number of arc seconds in a radian is 206265.
If we adopt {\em our} value of the radius of the Earth from \S 3 (namely, 6290 km),
we obtain a distance to the asteroid of (4.66 $\pm$ 0.56) $\times$ 10$^7$ km.  

\parindent = 9 mm

Using the Steeger orbit solution in Table \ref{orbit_det},
the method of \citet[][p. 125 ff.]{Mee88}, and the rectangular
coordinates of the Sun on 2008 July 24 from the {\em Astronomical
Ephemeris}, we determined that asteroid 1996 HW1 was 0.294 AU distant
when Figs. \ref{pos3} and \ref{pos4} were taken.  Our resulting value for 
the Astronomical Unit is (1.59 $\pm$ 0.19) $\times$ 10$^8$ km, which
is roughly 6 percent larger than the true value of 1.496 $\times$ 10$^8$ km.

Our value of the solar parallax, P$_S$ in Fig. \ref{lunar_ecl}, equals 
(6290/1.59 $\times$ 10$^8$) $\times$ 206265 = 8\farcs 2, or about
0\farcm 14, which we used in \S 4.1.  Our distance to the Sun
in Earth radii divided by our distance to the Moon in Earth radii
gives a Sun distance that is roughly 406 times the distance to
the Moon, which is close to the true mean value of 389.

\section {Systematic and Random Errors}

In the interests of readability we have not included all statistical
details above.  However, given the readership of the {\em American Journal of
Physics}, we thought it prudent to have a discussion of various kinds
of errors here.

Say you want to answer a simple question such as ``How tall am I?''  By
standing up against the wall, placing a ruler level on the top of your
head and making a mark on the wall you can find out the answer.  But
what if you forgot to take your shoes off?  You would have a {\em
systematic error} equal to the height of the heels of your shoes.

What if you measured your height immediately after getting out of bed
after a full night's sleep?  You would find that first thing in the
morning you are about 2 cm taller than you will be later that same day.
Why is this so?  The disks between the vertebrae in your back expand
when you are lying down sleeping.  Thus, if you measured your height 
first thing in the morning, right after lunch, then after a 6 mile
run in the afternoon, you could get three different answers.  The
``mean value of your height during the day'' might be 178 $\pm$ 1 cm. 
This standard deviation of $\pm$ 1 cm is the ``mean error of the mean'' 
of the set of measurements.  This is referred to as a {\em random error} because
values that you measure would scatter about some average value.

Very often a set of measurements exhibits a Gaussian (i.e. 
bell-shaped) distribution.  Then 68.3 percent of the 
measurements are within one standard deviation of the mean 
value, 95.5 percent are within 2-$\sigma$ and 99.7 percent are 
within 3-$\sigma$. This allows us to identify {\em outliers} in 
the data.  If some data value is several (or many) standard 
deviations from the mean (or expected) value, we have either 
underestimated the size of our random errors, or there is some 
unaccounted source of systematic error.

In \S2 above we derived a value of the latitude for a particular location in College
Station, Texas, which was 11\farcm~5 north of the true value.  This
is the systematic error.  Usually in science we do not know the ``true''
value, but in this case we do.  From a scatter of the measurements of
the shadow length of our gnomon we could obtain a numerical value of
the root-mean-square scatter of the data about some best fit line.
That gives us an estimate of the uncertainty of the minimum shadow
length, which translates into a random error for the maximum elevation
angle of the Sun on some given day, which then leads to a random error 
for our value of the latitude.  Our value of the latitude is within
3 of our standard deviations of the true value.  Our value of the
tilt of the Earth's axis of rotation to the plane of its orbit is 6
standard deviations from the true value.  Thus, we have room for
improvement in the experiment, should we choose to do it again and
again.  

Some sources of uncertainty have not yet been mentioned.  For the
location of the June 21st and December 21st gnomon experiments we
have measured the levelness of the spot where the data were taken,
finding that a perfectly straight gnomon which was perfectly squarely
set in its base would have been tilted
6 $\pm$ 1 arc minute north of the zenith.  This would have made the
elevation angle too low by that amount at local noontime, but
less so at other times.  To complicate matters, our gnomon is
just a wooden dowel rod which is {\em not} totally straight, and we
do not know just how squarely it sits in the block of wood that
is its base.  

Regarding the use of the ``map tool'' in \S3, we found that the 
great circle arc from South Bend, Indiana, to College Station, Texas, 
was 0.7437 $\pm$ 0.0009 of the route driven.  We have rounded this
off to 0.744 and have decided to ignore the numerical uncertainty.

In \S4.1 we found from six images of the Moon in partial lunar 
eclipse taken from 21:29:55 to 21:52:59 UT that the angular 
diameter of the Earth's shadow was 2.46, 2.65, 2.49, 2.62, 2.56, 
and 2.56 times the angular diameter of the Moon.  These were 
obtained using hard copies of the images, a compass, and a 
ruler.  In all instances we could see more than a 180 degree arc 
of the illuminated part of the Moon.  The mean ratio was 2.56 
$\pm$ 0.03. To find angle $\tau$ (the angular radius of the 
Earth's shadow) we used the (more robust) mean value of the 
angular diameter of the Moon from Krisciunas (2010) rather than 
an estimated value of the Moon's angular size on 2011 June 15.

In \S4.2 and Table \ref{separations} we give results of an 
experiment that is at or beyond the capabilities of a ruler plus 
cardboard cross piece. We found that the systematic error of the 
angular separation of two bright point sources (Saturn vs. 
$\alpha$ Vir) was +0.12 deg.  The random error was $\pm$0.04 
deg. These are considerably better than the often quoted 
uncertainty of 0.20 to 0.25 deg for non-telescopic measurements.  
But consider our measurements of the angular separation of 
Saturn vs. the Moon and $\alpha$ Vir vs. the Moon.  Our mean
systematic errors are as great as 0.39 deg and our random errors 
are $\pm$ 0.23 to $\pm$ 0.29 deg. Table \ref{posns} shows that if 
we have measured the Moon's position with respect to three or 
more point sources, we can locate the Moon's location on the sky 
to within 0.3 deg on average.  But if we have measured the 
Moon's position with respect to only two point sources, the 
error in position can be too large to be useful for determining 
the rate of change of the Moon's position over the course of a 
night.

When it comes to certain kinds of analysis, there are other
errors that are neither systematic or random.   In the top half
of Fig. \ref{shadow} we have fitted a hyperbola to the full
data set.  But this is for illustrative purposes only.  Why?
Careful scrutiny shows that the points do not randomly scatter
above and below the curved line.  The actual function we need
requires the latitude of the site, which is the thing we are
trying to {\em derive}.  In this case, for practical reasons
we have fitted a fourth order polynomial to a subset of the data
around the time of the minimum shadow length.  The bottom line
is that we fit some data under the {\em assumption}
that we have the appropriate function.  What the 
most appropriate function is also depends on the typical size of the
error bars of the data points.  If the error bars
are large, high order polynomial fits would be unjustified.

\section{Discussion and Conclusions}

Using a vertical stick, a car, watch, map, and map tool, we have
measured the size of the Earth from first principles.  We used
observations of the elevation angle of the Sun over the course
of a few hours overlapping local noontime on the summer solstice
and winter solstice of 2010 to determine the latitude and
longitude of College Station, Texas.  We also determined the
latitude and longitude of South Bend, Indiana.  The
two sites were 13.78 deg apart along a great circle arc.  We
obtained a value of the radius of the Earth of 6290 km, 
about 1.4 percent too small.

Aristarchus's method of determining the distance to the Moon derives
from the geometry of the Earth, Moon, and Sun at the time of a lunar
eclipse.  One key observable is the angular size of the Earth's
shadow at the distance of the Moon compared to the angular size of the 
Moon.  Using images available on the web of the lunar eclipse
of 2011 June 15 we found that the Earth's
shadow was 2.56 times the angular diameter of the Moon.  The corresponding
distance of the Moon was 62.3 Earth radii, about 3.3 percent larger
than the true mean value.

It is also possible to determine the distance to the Moon from observations
of its motion against the background of stars.  These observations are
made from a single site over the course of one night.  This method is
far more complicated than Aristarchus's method, but from such observations
we showed that the Moon was between 47 and 62 R$_{\earth}$ distant.  
The true range of the Moon's distance is 55.9 to 63.8 R$_{\earth}$.

Simultaneous observations from Socorro, New
Mexico, and Ojai, California allowed us to determine the distance
to the asteroid 1996 HW1 on 2008 July 24.  A determination of the 
orbital elements of this asteroid by one of us allowed us to 
calculate that the asteroid was 0.294 AU distant on that date.  
The final result was a calibration of the AU equal to 
(1.59 $\pm$ 0.19) $\times$ 10$^8$ km, which is 6 percent larger 
than the true value.  

Given the seeing and tracking constraints associated with our asteroid
observations, our asteroid experiment was at the limits of the technology
of our small telescopes.  To determine the distance to a {\em main belt}
asteroid (which would have been 10 times more distant) would have required
a baseline 10 times bigger than the 1130 km baseline we used if we
wanted the parallax to be several arc seconds.  Such
an experiment would be nearly impossible on the surface of the Earth.

This paper has overflowed with numerical and trigonometrical details.
Still, the {\em qualitative} results are easy to understand and are worth restating.  
Using very inexpensive equipment we can determine the size of the Earth 
and the distance to the Moon.  To determine the length of the AU
we do {\em not} need to organize an international endeavor like what
was done for the Venus transits of 1761, 1769, 1874, and 1882 \citep{Van95}.
We can measure the length of the AU with carefully timed observations
of a near Earth asteroid using telescopes comparable to those owned
by many amateur astronomers.

It is also worth reminding ourselves why astronomers have such an obsession
with determining cosmic distances.  Thanks to Kepler's Third Law
the length of the AU gets us the distances to all other objects that
orbit the Sun.  Once we have the distances to stars like the Sun
(converting hydrogen to helium in their cores, but with a range of
mass) via trigonometric stellar parallaxes -- the next rung of the distance
ladder -- we can use photometric methods to determine distances to
star clusters using the fact that ``main sequence stars'' of the
same mass have comparable intrinsic brightness.  A simple equation
relates the apparent brightness of a star with its intrinsic
brightness and the distance.  We can 
exploit the Cepheid period-luminosity law to calibrate our way 
across the Galaxy and to nearby galaxies.  Knowing the energy budget
of a star helps us determine what it is made of, what is its
structure, how it will live, and how it will die.  Luminosities of 
Type Ia supernovae have allowed us to address some the largest
questions we can ask, such as, ``What is the ultimate fate of the Universe?''

The interconnectedness of astronomical topics means that the
big questions are related to the Earth size experiment of 
Eratosthenes carried out more than 22 centuries ago.  Starting with
an instrument as simple as a vertical stick we can connect 
basic and profound aspects of the universe we inhabit.

\vspace {1 cm}

\acknowledgments

Tom Weimar helped with the construction of the gnomon pictured in Fig. \ref{gnomon}.
We thank Nick Suntzeff for useful discussions.
The NMT observations of asteroid 1996 HW1 were obtained with the help of
Jeff Lu, David Oh, Anna Heinz, Peter Combs, and Rebecca Mickol.  We thank 
Dan Klinglesmith and Jason Speights for observing support at Etscorn 
Observatory.  The asteroid observations were made as part of 
the Summer Science Program, which is funded by private donations
and sponsored by New Mexico Institute of Mining and Technology.
Elisabeth Button kindly made Fig. 1.

\newpage

\begin{deluxetable}{clcl}
\tabletypesize{\scriptsize}
\tablewidth{0pc}
\tablecaption{Gnomon Data\tablenotemark{a}\label{gnomon_data}}
\tablehead{   \colhead{CDT (Jun 21)} & \colhead{L (mm)} 
&\colhead{CST (Dec 21)} & \colhead{L (mm)}  } 
\startdata
11:25:00 & 330       & 11:07:45 & 844   \\
11:35:00 & 300.5     & 11:15:00 & 829   \\
11:46:30 & 268.5     & 11:25:00 & 807   \\
11:55:00 & 245.5     & 11:35:15 & 787   \\
12:06:17 & 217       & 11:49:20 & 770   \\
12:15:04 & 196       & 11:59:20 & 761.5 \\
12:25:02 & 172       & 12:10:00 & 756.5 \\
12:35:01 & 151       & 12:15:00 & 754.5 \\
12:45:06 & 131       & 12:20:00 & 754   \\
12:55:00 & 112.5     & 12:25:15 & 754   \\
13:00:10 & 104       & 12:30:00 & 755   \\
13:05:00 & \phs 99   & 12:36:00 & 756.5 \\
13:10:47 & \phs 92.5 & 12:40:38 & 759.5 \\
13:15:00 & \phs 88.5 & 12:45:18 & 761.5 \\
13:20:00 & \phs 86.5 & 12:50:00 & 765.5 \\
13:25:00 & \phs 86   & 12:55:43 & 771   \\
13:30:10 & \phs 86   & 13:00:00 & 775   \\
13:37:20 & \phs 92   & 13:05:20 & 782   \\
13:45:00 & 101       & 13:10:06 & 788.5 \\ 
13:49:30 & 106       & 13:20:00 & 805   \\
13:56:00 & 117       & 13:30:00 & 824.5 \\
14:04:30 & 132       & 13:40:00 & 848   \\
14:15:00 & 154       & 13:50:00 & 875   \\
14:25:00 & 175.5     & 14:01:20 & 910   \\
14:38:12 & 206       & 14:10:50 & 946   \\
14:45:00 & 224       & 14:20:06 & 987   \\
14:55:00 & 247.5     &          &       \\
15:05:00 & 275       &          &       \\
15:15:00 & 303       &          &       \\
15:25:00 & 331.5     &          &       \\
\enddata
\tablenotetext{a}{Gnomon height was 632 $\pm$ 1 mm for the 2010 June 21
observations, and 550 $\pm$ 1 mm for the 2010 December 21  observations.  
Column 1 is Central Daylight Time.  Columns 2 and 4 gives the shadow lengths in mm.
Column 3 is Central Standard Time. }
\end{deluxetable}

\begin{deluxetable}{clrrl}
\tablewidth{0pc}
\tablecaption{Angular separations (2011 May 15)\label{separations}}
\tablehead{   \colhead{UT} & \colhead{Object pair} & 
\colhead{$\rho _{obs}$} & \colhead{$\rho _{true}$} & \colhead{diff\tablenotemark{a}}  }
\startdata
02:21:00  & $\alpha$ Vir vs. Moon & 3.\hspace{-0.8 mm}$^{\circ}$41  & 3.\hspace{-0.8 mm}$^{\circ}$70  &   +0.\hspace{-0.8 mm}$^{\circ}$29 \\
02:24:00  & Saturn vs. Moon       & 12.85 & 13.21 &   +0.36 \\
02:28:00  & $\delta$ Crv vs. Moon & 10.71 & 10.85 &   +0.14 \\
02:30:00  & Saturn vs. $\alpha$ Vir & 13.60 & 13.71 & +0.11 \\
\hline
04:04:00  & $\alpha$ Vir vs. Moon & 3.28  & 3.40  &   +0.12 \\
04:06:40  & Saturn vs. Moon       & 13.34 & 13.86 &   +0.52 \\
04:08:00  & $\delta$ Crv vs. Moon & 11.33 & 11.34 &   +0.01 \\
04:10:00  & Saturn vs. $\alpha$ Vir & 13.63 & 13.71 & +0.08 \\
\hline
05:54:00  & $\alpha$ Vir vs. Moon & 3.32  & 3.22  & $-$0.10 \\
05:56:00  & Saturn vs. Moon       & 14.52 & 14.53 &   +0.01 \\
05:58:30  & $\delta$ Crv vs. Moon & 11.77 & 11.91 &   +0.14 \\
06:01:40  & Saturn vs. $\alpha$ Vir & 13.55 & 13.71 & +0.16 \\
06:04:00  & $\gamma$ Hya vs. Moon & 9.33  & 9.31  & $-$0.02 \\
\hline
08:02:00  & $\alpha$ Vir vs. Moon & 3.63  & 3.15  & $-$0.48 \\
08:03:00  & Saturn vs. Moon       & 14.80 & 15.34 &   +0.54 \\
08:05:00  & Saturn vs. $\alpha$ Vir & 13.55 & 13.71 & +0.16 \\
\hline
08:31:00  & Saturn vs. $\alpha$ Vir & 13.60 & 13.71 & +0.11 \\
08:38:00  & $\alpha$ Vir vs. Moon & 3.35  & 3.17  & $-$0.18 \\
08:41:00  & Saturn vs. Moon       & 15.06 & 15.60 &   +0.54 \\
\enddata
\tablenotetext{a}{In the sense ``true'' minus ``observed''.}
\end{deluxetable}

\begin{deluxetable}{cccccc}
\tablewidth{0pc}
\tablecaption{Derived and true topocentric positions of Moon (2011 May 15)\label{posns}}
\tablehead{\colhead{UT} & \colhead{$\alpha _{obs}$} & 
\colhead{$\delta _{obs}$} & \colhead{$\alpha _{true}$} & \colhead{$\delta _{true}$} &
\colhead{$\rho$\tablenotemark{a}} }
\startdata
02:24:00 &  198.\hspace{-0.8 mm}$^{\circ}$15 $\pm$ 0.09 & $-$12.\hspace{-0.8 mm}$^{\circ}$76 $\pm$ 0.09 & 
198.\hspace{-0.8 mm}$^{\circ}$27 & $-$13.\hspace{-0.8 mm}$^{\circ}$20 & 0.\hspace{-0.8 mm}$^{\circ}$46 \\
04:04:40 &  198.76 $\pm$ 0.04 & $-$13.14 $\pm$ 0.07 & 198.93 & $-$13.56 & 0.45 \\
05:56:00 &  199.53 $\pm$ 0.02 & $-$13.94 $\pm$ 0.01 & 199.63 & $-$13.90 & 0.10 \\
08:03:00 &  199.50 $\pm$ 0.57 & $-$14.31 $\pm$ 0.36 & 200.55 & $-$14.25 & 1.02 \\
08:41:00 &  200.03 $\pm$ 0.62 & $-$14.27 $\pm$ 0.34 & 200.87 & $-$14.34 & 0.81 \\
\enddata
\tablenotetext{a}{Angular distance between derived position of Moon based on 
observations with cross staff vs. true topocentric position of Moon.  For the
fourth and fifth determinations the uncertainties in the right ascension
and declination derive from the assumption that the angular separations of the Moon
vs. $\alpha$ Vir and the Moon vs. Saturn were accurate to $\pm$ 0.3 deg.}
\end{deluxetable}

\begin{deluxetable}{clll}
\tablewidth{0pc}
\tablecaption{Steeger orbit determination for asteroid 1996 HW1\label{orbit_det}}
\tablehead{\colhead{Parameter} & \colhead{Description} & \colhead{Steeger value} & \colhead{JPL Horizons value}} 
\startdata
t$_0$       &    epoch (UT)            & 2008 June 29.35             &  2007 January 15.00 \\
t$_0$       &    epoch (Julian Date)   & 2454646.85                  &  2454115.5          \\
$M$         &    mean anomaly at t$_0$ & 335.\hspace{-0.8 mm}$^{\circ}$1461 & 155.\hspace{-0.8 mm}$^{\circ}$33678\tablenotemark{a}  \\
$a$         &    semi-major axis       & 2.0855 AU                          & 2.046041 AU \\
$e$         &    eccentricity          & 0.4575                             & 0.449165   \\
$i$         &    inclination angle     & 8.\hspace{-0.8 mm}$^{\circ}$5033   & 8.\hspace{-0.8 mm}$^{\circ}$437363     \\
$\Omega$    &    longitude of ascending node & 177.\hspace{-0.8 mm}$^{\circ}$6887  & 177.\hspace{-0.8 mm}$^{\circ}$216737 \\
$\omega$    &    argument of perihelion & 176.\hspace{-0.8 mm}$^{\circ}$1618   & 177.\hspace{-0.8 mm}$^{\circ}$020070\\
\enddata
\tablenotetext{a}{The mean anomaly increases by 0.33676929 deg per day according
to JPL Horizons, so it is 334.\hspace{-0.8 mm}$^{\circ}$2791 at
the epoch of the Steeger solution.}
\end{deluxetable}

\clearpage

\figcaption[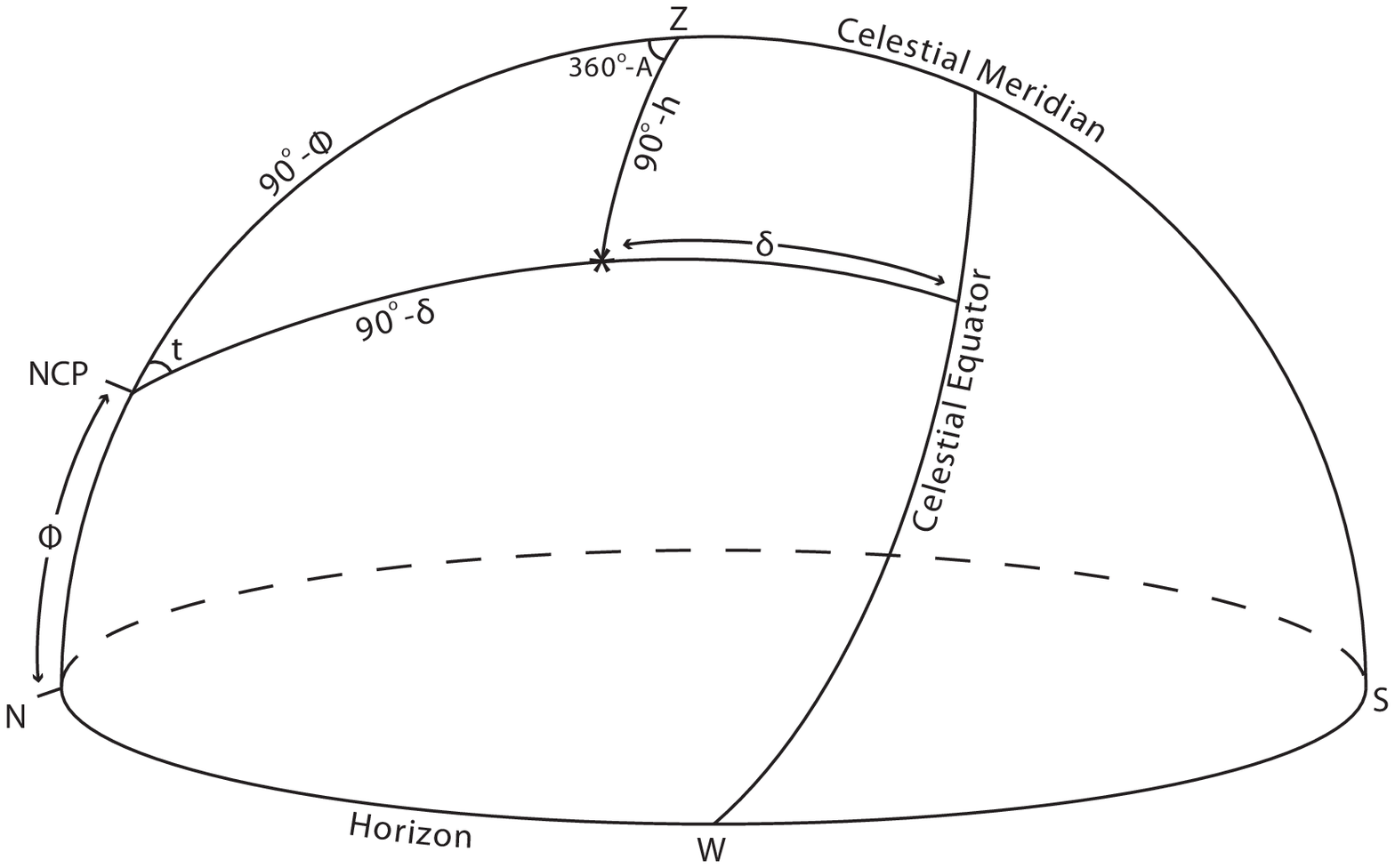]
{The astronomical triangle. The North Celestial Pole is labeled ``NCP".  
The zenith is at Z.   The latitude of the site is $\phi$.  An object
in the western sky is marked by an asterisk.  The hour angle of the object is $t$,
which is positive in the western sky, negative in the eastern sky.  The declination
of the object is $\delta$.  Its elevation angle above the horizon is $h$, so
the zenith angle is 90$^{\circ} - h$.  
The azimuth A of an object is measured clockwise around the horizon, with north = 0$^{\circ}$.
\label{triangle}
}

\figcaption[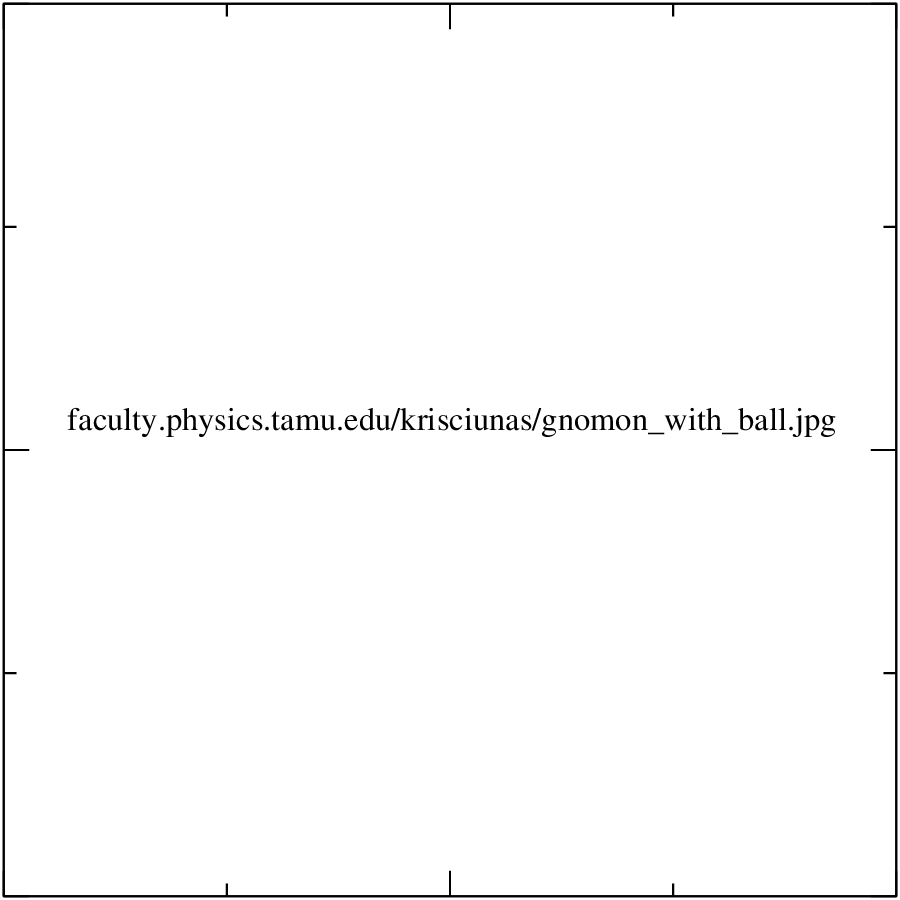]
{A gnomon.  It consists of a wooden base with a hole drilled through it
using a drill press, and a vertical stick that fits tightly.  It can
be a pointed stick.  Here we have fashioned a small sphere at the top.  It
is easier to measure the center of the elliptical shadow of the sphere than
the end of the darker part of the shadow of a pointed stick.
\label{gnomon}
}

\figcaption[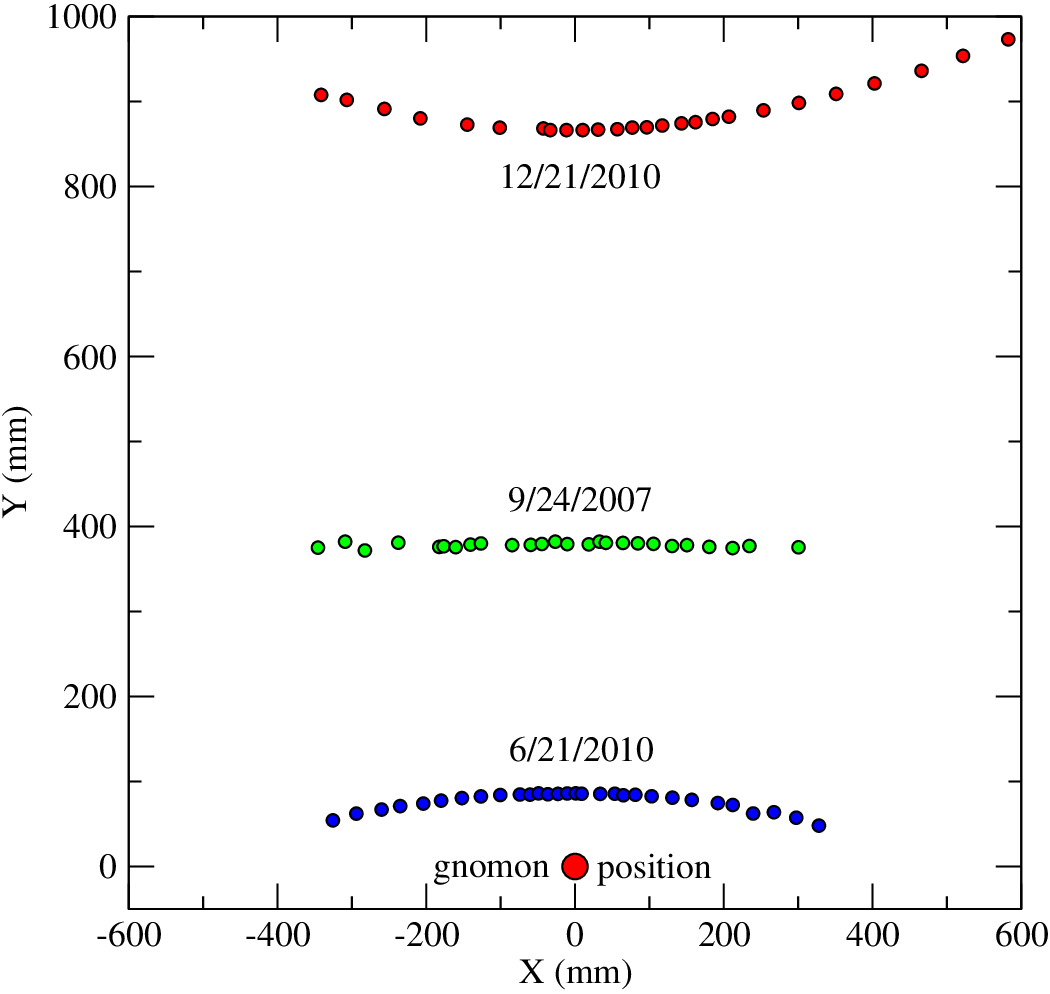]
{The X-Y positions of the end of the shadow of a 632 mm high gnomon
used in College Station, Texas.  For the 2010 December 21 observations
we have scaled the coordinates by 632/550, as a 550 mm high gnomon was used
on that date.  The curvature of these loci change with the declination
of the Sun over the course of the year.  In fact, the shadow lengths and the 
X-positions, along with the hour angle of the Sun obtained from the
times of the observations and the time of minimum shadow length, allow
us to derive the declination of the Sun.  In that case it is best to use 
observations obtained when the Sun is roughly one hour or more from the meridian.
\label{gnomon_xy}
}

\figcaption[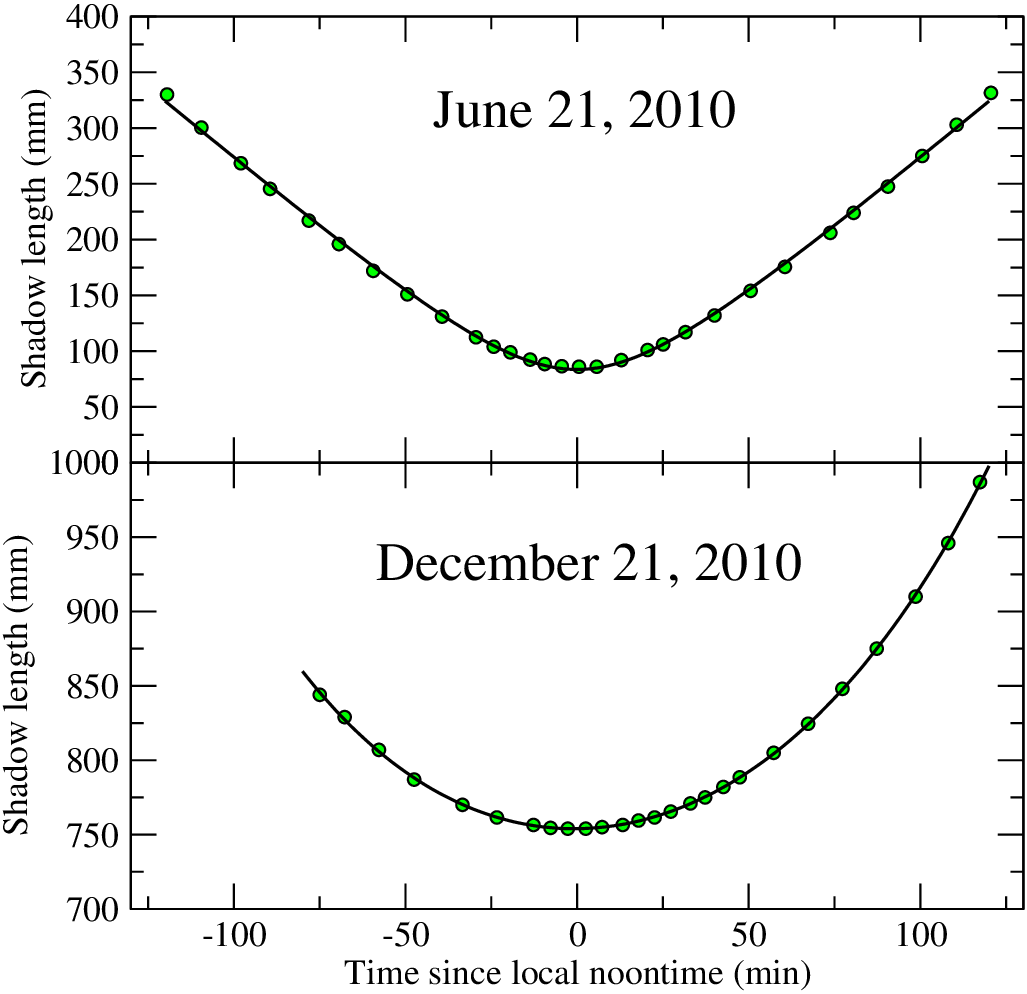]
{Upper figure: the shadow length of a 632 mm gnomon on the day of the
summer solstice, as measured in College Station, Texas.  For illustrative
purposes {\em only} a hyperbola is fit to the data.  Lower figure: the shadow
length of a 550 mm gnomon on the day of the winter solstice, as measured
at the exact same location.  A fourth order polynomial is fit to the data.
\label{shadow}
}

\figcaption[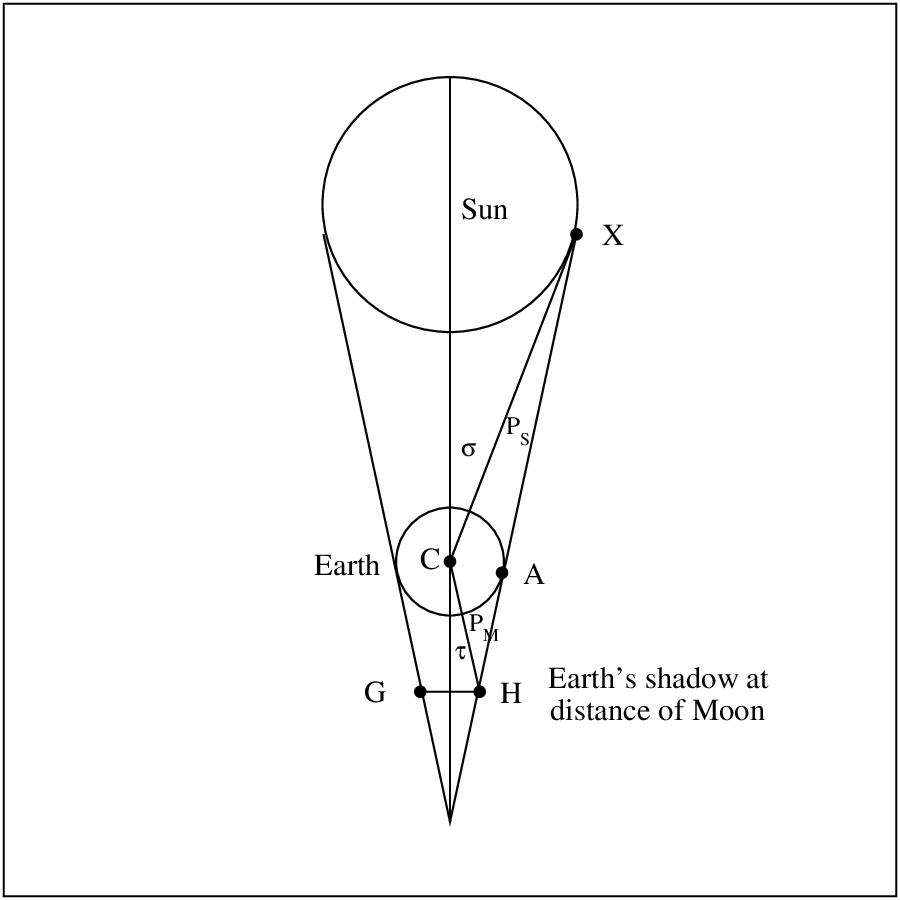]
{Geometry of the lunar eclipse (not to scale).  At point G the
Moon is halfway into the shadow of the Earth.  At point
H the Moon is halfway out of the shadow. Simple arguments 
and measurements originating with Aristarchus allow us to estimate 
the distance to the Moon in terms of the Earth's radius.
\label{lunar_ecl}
}

\figcaption[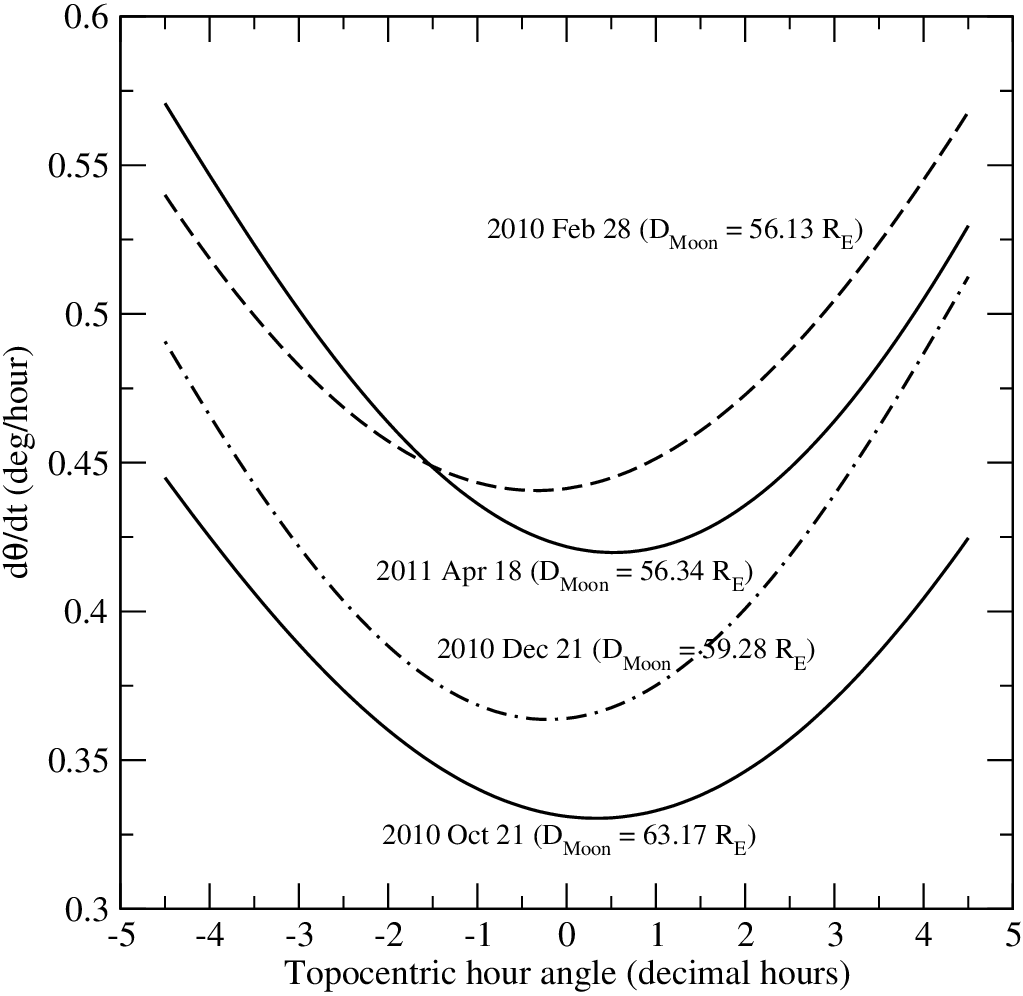]
{The angular motion of the Moon as a function of hour angle for
an observer in College Station, Texas.
\label{moon_rate}
}

\figcaption[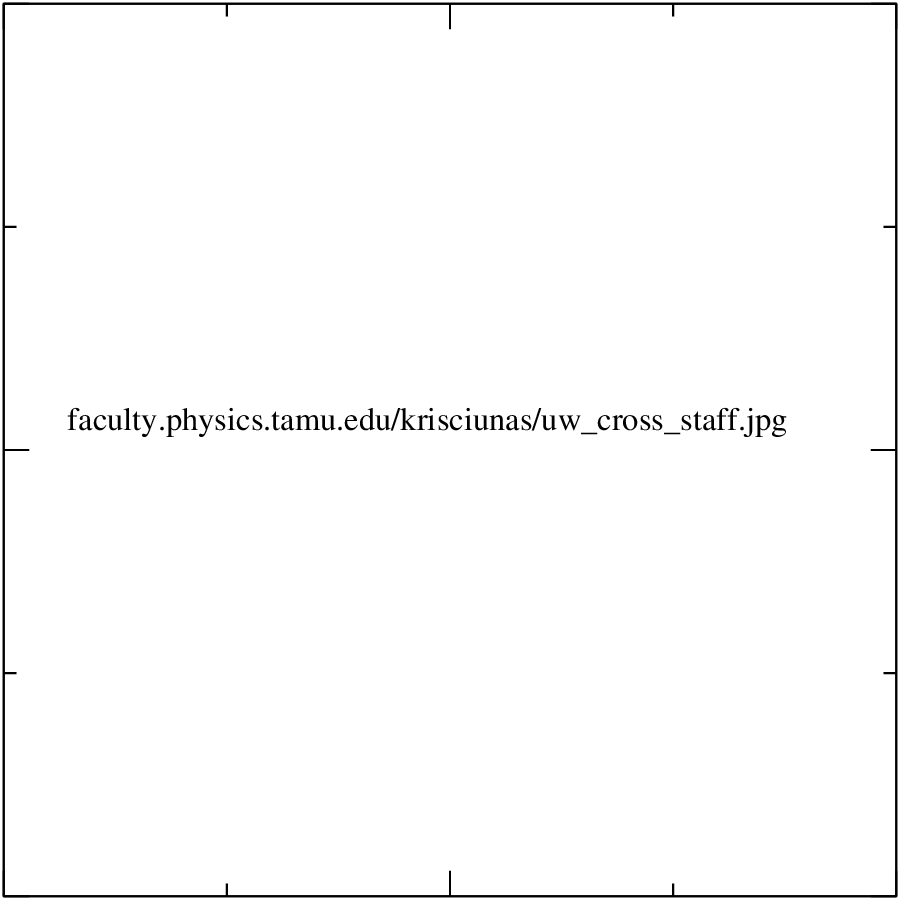]
{The cross staff.  The cardboard cross piece slides up and down the
yardstick.  Using simple geometry we can use this device to determine
the angular separation of objects in the sky.
\label{cross_staff}
}

\figcaption[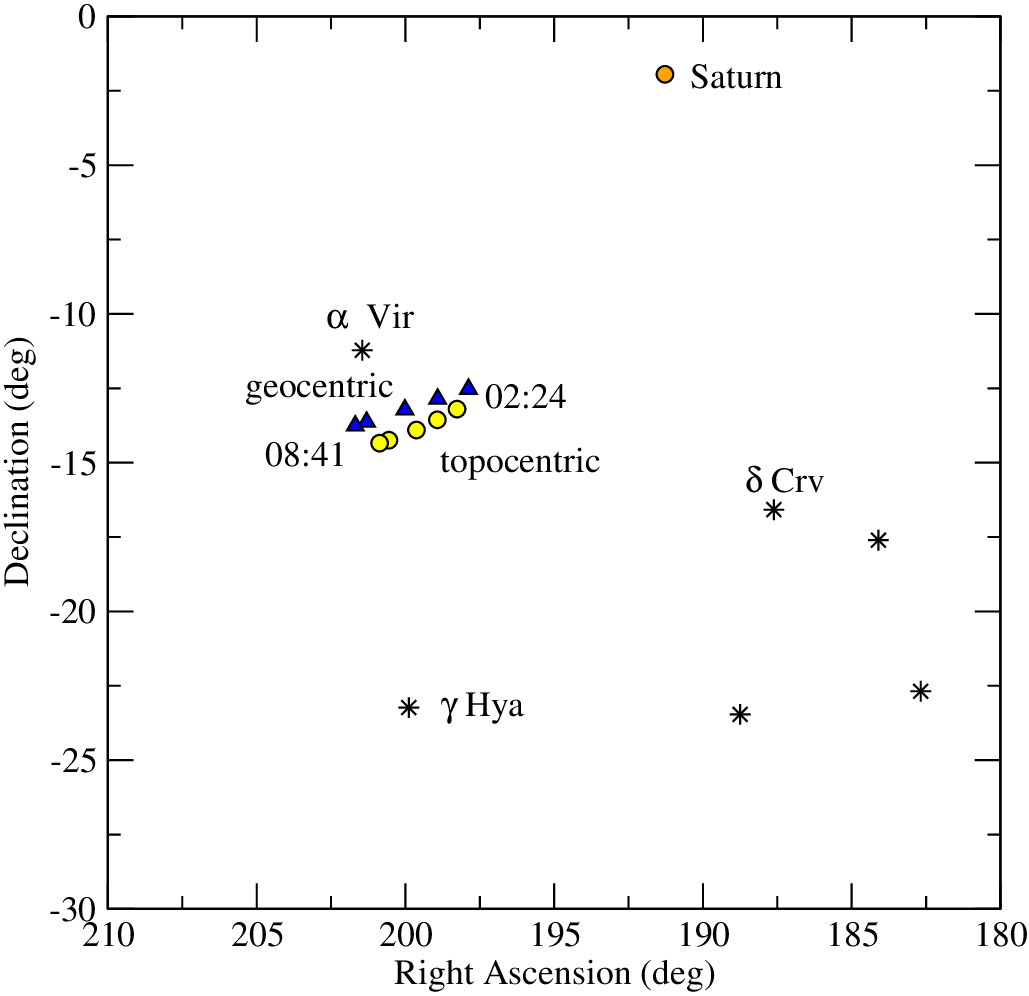]
{The topocentric and geocentric RA and DEC of the Moon on 2011 May 15.
The topocentric values are calculated for College Station, Texas.
The zenith angle of the Moon is greater for an observer situated
on the surface of the Earth compared to a hypothetical observer at
the center of the Earth.  In other words, by observing on the surface
of the Earth, the Moon appears to be lower in the sky compared to
what would be seen by a hypothetical observer at the center of 
the Earth.  This elevation angle offset translates to varying 
shifts in RA and DEC over the course of the night.
\label{may15}
}

\figcaption[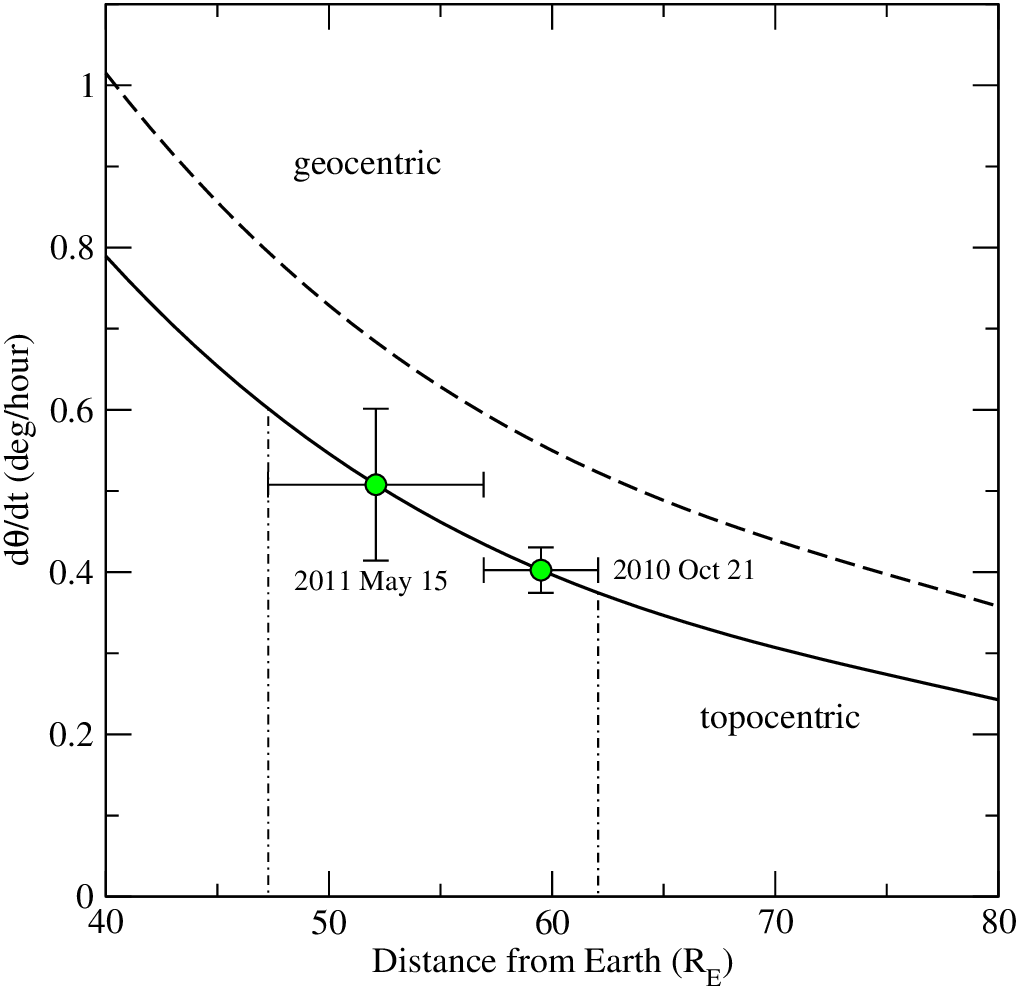]
{Dashed curve: angular motion of an Earth-orbiting object 
as viewed by a hypothetical observer at the Earth's center, as a 
function of its distance in R$_{\earth}$.  Solid curve:
mean angular motion of an Earth-orbiting object as viewed from
College Station, Texas.  The average is taken over an 8 hour period
centered on the meridian transit.  
On 2010 October 21 we measured the Moon to move 0.40 $\pm$ 0.03 
deg/hr from six observations over 9.1 hours.  The implied value of the
Moon's distance is between 57 and 62 R$_{\earth}$ on that occasion.
On 2011 May 15 we measured the Moon to move 0.508 $\pm$ 0.094 deg/hr
from three sets of observations taken over 3.9 hours.  The implied 
distance to the Moon is roughly 47 to 58 R$_{\earth}$.
\label{moon_obs}
}

\figcaption[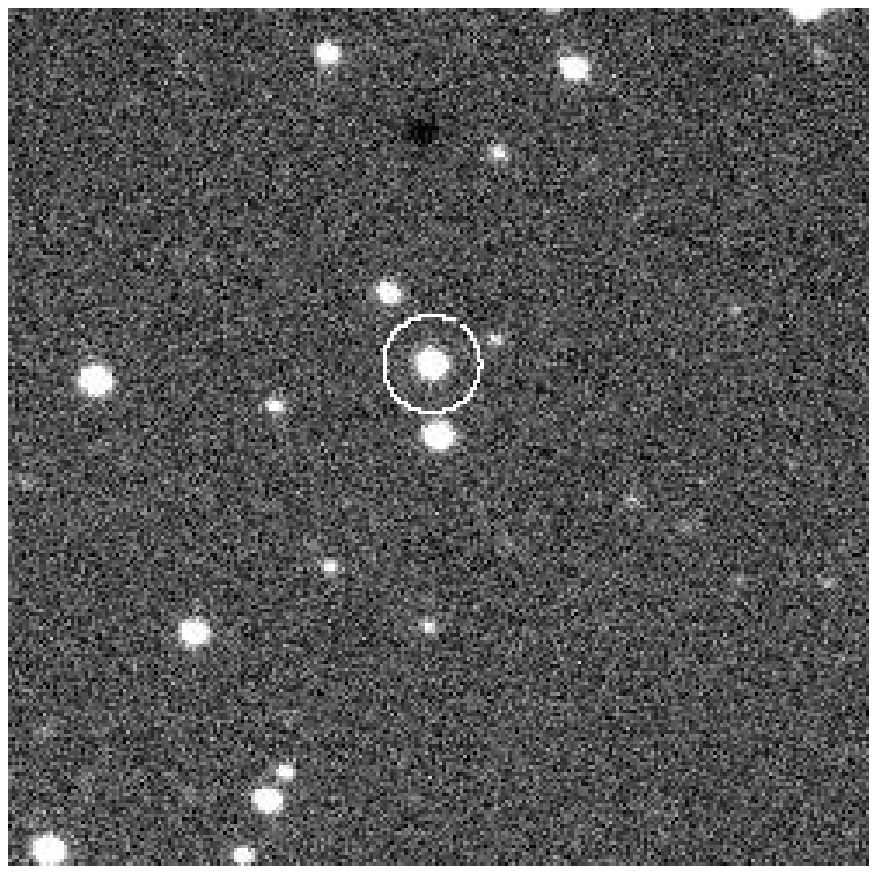]
{Asteroid 1996 HW1 is circled in this unfiltered 90 second image obtained 
by E. DeBenedictis with a 15 cm Takahashi refractor at Etscorn
Observatory, Socorro, New Mexico, on 2008 July 24 at 08:17:27.8 UT. 
North is up, east is to the left.
\label{pos3}
}

\figcaption[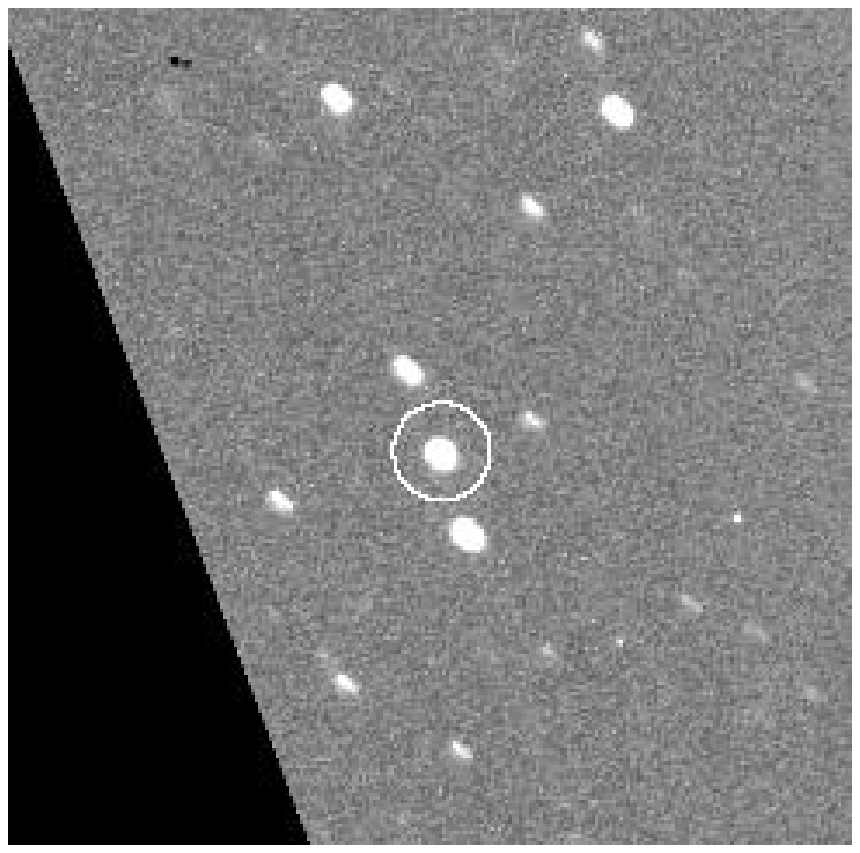]
{Asteroid 1996 HW1 is circled in this unfiltered 90 second image obtained 
by G. Tabak and K. Pasricha with a 25 cm Meade reflector at Besant Hill
School in Ojai, California, on 2008 July 24 at 08:17:28 UT, with an
uncertainty of no more than $\pm$ 2 seconds.  North is up, east to the left.
Note that this image was taken at the same time as Fig. \ref{pos3}, but
the asteroid is roughly 5 arc seconds to the left (east) as observed
at the other end of a baseline of 1130 km.
\label{pos4}
}

\clearpage

\begin{figure}
\plotone{f1.eps}
{\center Krisciunas {\it et al.} Fig. \ref{triangle}}
\end{figure}

\begin{figure}
\plotone{f2.eps}
{\center Krisciunas {\it et al.} Fig. \ref{gnomon}}
\end{figure}

\begin{figure}
\plotone{f3.eps}
{\center Krisciunas {\it et al.} Fig. \ref{gnomon_xy}}
\end{figure}

\begin{figure}
\plotone{f4.eps}
{\center Krisciunas {\it et al.} Fig. \ref{shadow}}
\end{figure}

\begin{figure}
\plotone{f5.eps}
{\center Krisciunas {\it et al.} Fig. \ref{lunar_ecl}}
\end{figure}

\begin{figure}
\plotone{f6.eps}
{\center Krisciunas {\it et al.} Fig. \ref{moon_rate}}
\end{figure}

\begin{figure}
\plotone{f7.eps}
{\center Krisciunas {\it et al.} Fig. \ref{cross_staff}}
\end{figure}

\begin{figure}
\plotone{f8.eps}
{\center Krisciunas {\it et al.} Fig. \ref{may15}}
\end{figure}

\begin{figure}
\plotone{f9.eps}
{\center Krisciunas {\it et al.} Fig. \ref{moon_obs}}
\end{figure}

\begin{figure}
\plotone{f10.eps}
{\center Krisciunas {\it et al.} Fig. \ref{pos3}}
\end{figure}

\begin{figure}
\plotone{f11.eps}
{\center Krisciunas {\it et al.} Fig. \ref{pos4}}
\end{figure}

\end{document}